\documentclass[12pt,preprint,psfig]{aastex}

\usepackage{emulateapj5}

\def\ltsima{$\; \buildrel < \over \sim \;$}

\def\lsim{\lower.5ex\hbox{\ltsima}}
\def\gtsima{$\; \buildrel > \over \sim \;$}
\def\gsim{\lower.5ex\hbox{\gtsima}}

\begin{document}
\title{Improved Constraints on The Neutral Intergalactic Hydrogen
Surrounding Quasars at Redshifts $z>6$}

\author{J. Stuart B. Wyithe\altaffilmark{1}, Abraham Loeb\altaffilmark{2},
Chris Carilli\altaffilmark{3}}

\email{swyithe@isis.ph.unimelb.edu.au; aloeb@cfa.harvard.edu; ccarilli@aoc.nrao.edu}

\altaffiltext{1}{University of Melbourne, Parkville, Victoria, Australia}

\altaffiltext{2}{Astronomy department, Harvard University, 60 Garden
St., Cambridge, MA 02138}

\altaffiltext{3}{National Radio Astronomy Observatory, P. O. Box 0,
Socorro, NM 87801}

\begin{abstract}
\noindent 
We analyze the evolution of \ion{H}{2} regions around the seven known SDSS
quasars at $z>6$. The comparison between observed and model radii of the
\ion{H}{2} regions generated by these quasars individually, suggests that
the surrounding intergalactic hydrogen is significantly neutral.  When
all constraints are combined, the existing quasar sample implies a
volume averaged neutral fraction that is larger than 10\% at
$z>6$. This limited sample permits a preliminary analysis of the
correlations between the quasar parameters, the sizes of their
\ion{H}{2} regions, and the associated constraints on the neutral
hydrogen fraction.  We find no evidence in these correlations to
contradict the interpretation that the red side of the Gunn-Peterson
trough corresponds to the boundary between an \ion{H}{2} region and a
partially neutral IGM.

\end{abstract}


\section{Introduction}

The Gunn-Peterson (1965, hereafter GP) troughs in the spectra of the most
distant quasars at redshifts $z\sim 6.3$--$6.4$ (Fan et al.~2004) hint that
the reionization of cosmic hydrogen was completed only a billion years
after the big bang.  Unfortunately, the troughs only set a lower limit of
$\sim 10^{-3}$ on the volume averaged neutral fraction (Fan et al.~2001;
White et al. 2003). This leaves open the question of whether reionization
was completed only at $z\sim6$ so that the GP trough is due to a
significantly neutral IGM, or whether it is only a residual neutral
fraction left over from reionization at an earlier epoch that is
responsible for the observed Ly$\alpha$ absorption. This question has
become particularly acute following evidence presented by the WMAP
satellite in favor of early reionization (Kogut et al.~2003; Spergel et
al.~2003). Indeed, if reionization ended only at $z\sim6$, then more exotic
``double reionization'' scenarios involving a massive early population of
stars may be required (Wyithe \& Loeb~2003a; Cen~2003).

The \ion{H}{2} regions generated by luminous quasars (Madau \&
Rees~2000; Cen \& Haiman~2000) expand faster into a medium with a
lower neutral fraction. Given information on the age of the quasar, we
can then use the observed size of the \ion{H}{2} regions to estimate
the neutral fraction. In an earlier paper we have shown that the size
of the \ion{H}{2} regions around the two highest redshift quasars
improves the lower limit obtained from the optical depth in the GP
trough by two orders of magnitude [Wyithe \& Loeb~(2004a); see also
supporting evidence by Messinger \& Haiman (2004) who analyze
SDSS~1030+0524 and find that the shape of the absorption spectrum near
the edge of the \ion{H}{2} region requires a large neutral fraction].
A large neutral fraction at $z>6$ is in contradiction with the
inference of a highly ionized IGM at $z\sim6.5$ which has been made
following the discovery of Ly$\alpha$ emitting galaxies (e.g. Rhoads
et al.~2004). However, the latter inference results from the assertion
that the red-damping wing of resonant Ly$\alpha$ absorption in the
surrounding IGM would lead to a strong suppression of the Ly$\alpha$
line emitted by these galaxies. This naive interpretation fails to
account for clustering of sources which can generate large \ion{H}{2}
regions around the Ly$\alpha$ emitters in an otherwise neutral IGM,
and therefore allow transmission of the Ly$\alpha$ line (Gnedin \&
Prada~2004; Furlanetto, Hernquist \& Zaldariaga~2004; Wyithe \&
Loeb~2004c).

In this paper, we supplement an earlier analysis of the \ion{H}{2} regions
around SDSS 1148+5251 and SDSS 1030+0524 (Wyithe \& Loeb~2004a) with new
data on six additional quasars at $z>6$. Our improved analysis
incorporates updated redshift data as well as the uncertainties in sizes of
the \ion{H}{2} regions, and in the ionizing luminosities of quasars. In
addition to the constraints derived from individual quasars, we compute
combined constraints on the neutral fraction using all quasars. We also
search for correlations between the quasar parameters, the sizes of their
\ion{H}{2} regions, and the associated constraints on the neutral fraction.

The paper is organized as follows. In \S~\ref{properties} we summarize
the observed properties of the sample, discuss the possibilities that
a short Ly$\alpha$ mean-free-path or dense absorbing cloud could mimic
signature of an \ion{H}{2} region, and consider the correlations
between different observables. In \S~\ref{model} we describe our model
for the evolution of an \ion{H}{2} region surrounding a luminous $z>6$
quasar. The limits that can be placed on the average neutral fraction
of hydrogen in the intergalactic medium (IGM) around individual
quasars are then described in \S~\ref{constraints}. Finally, we
summarize our conclusions in \S~\ref{conclusion}. Throughout the paper
we adopt the set of cosmological parameters determined by the {\em
Wilkinson Microwave Anisotropy Probe} (WMAP, Spergel et al. 2003),
namely mass density parameters of $\Omega_{m}=0.27$ in matter,
$\Omega_{b}=0.044$ in baryons, $\Omega_\Lambda=0.73$ in a cosmological
constant, and a Hubble constant of $H_0=71~{\rm
km\,s^{-1}\,Mpc^{-1}}$.

\clearpage

\section{Properties of the known $z > 6$ quasars}
\label{properties}

The Sloan Digital Sky Survey ({\it SDSS}) has discovered seven quasars at
$z \ge 6$ (Fan et al.  2001;2003;2004). These quasars are listed in
Table~\ref{tab1}. Column~2 lists the UV magnitudes. Column~3 lists the
source redshifts based on optically observed broad lines, such as
Ly$\alpha$ and high ionization lines such as \ion{N}{5}, \ion{C}{4},
\ion{Si}{4}. Column~4 lists redshifts based on lower ionization lines
observed in the near-IR, dominated by \ion{Mg}{2} but including the
\ion{Fe}{2} complex (Maiolino et al. 2004; Freudling et al. 2003; Iwamuro
et al. 2004; Willott, McClure, \& Jarvis 2003). Column~5 lists the CO
redshift for J1148+5251, and column~6 lists the redshift for the on-set of
the GP absorption (see below). Column~7 lists the difference between the
host galaxy redshift (see below) and the GP redshift, while column~8 gives
the corresponding size ($R$) in physical Mpc of the \ion{H}{2} region.

\begin{table*}[t]
\begin{center}
\caption{\label{tab1} Properties of $z>6$ quasars}
\begin{small}
\begin{tabular}{ccccccccc}
\hline 
Source & $M_{1450}$ & $z_{Ly\alpha,CIV}$ & $z_{MgII}$ & $z_{CO}$ & $z_{GP}$ & $z_{host} - z_{GP}$ & $\bar{R}\pm\sigma_R$ & $J^{\rm q}_{21}(R)$\\
\hline 

J1148+5251 & -27.82$^a$ & $6.37\pm0.03^c$ & $6.403\pm0.005^g$ & $6.419\pm0.001^b$ & 
$6.325\pm0.02^c$ & $0.09\pm0.02$ & 4.89$\pm$1.09 & 1.0\\ 

J1030+0524 & -27.15$^d$ & $6.28\pm0.03^c$ & $6.311\pm0.005^g$ & ~ & $6.178\pm0.005^c$ 
& $0.133\pm0.007$ & 7.50$\pm$0.39  & 0.2\\

J1048+4637 & -27.55$^a$ & $6.23\pm0.03^a$ & $6.203\pm0.005^g$ & ~ & $6.16\pm0.03^a$ & 
$0.04\pm0.03$ &   2.34$\pm$1.76  & 3.4 \\

J1623+3112 & -26.67$^e$ & $6.22\pm0.03^e$ &                & ~ & $6.16\pm0.03^e$ & 
$0.08\pm0.04$ & 4.65$\pm$1.74  & 0.4 \\

J1602+4228 & -26.82$^e$ & $6.07\pm0.03^e$ & ~ & ~ & $5.95\pm0.03^e$ & $0.12\pm0.04$ &   7.36$\pm$2.45  & 0.2 \\ 

J1630+4012 & -26.11$^a$ & $6.05\pm0.03^a$ & $6.065\pm0.005^g$ & ~ & $5.98\pm0.03^a$ & 
$0.085\pm0.03$ &   5.22$\pm$1.84  & 0.2 \\

J1306+0356 & -27.19$^d$ & $5.99\pm0.03^d$ & $5.99\pm0.02^f$ & ~ & $5.90\pm0.03^d$ & 
$0.09\pm0.04$ &   5.68$\pm$2.52  & 0.4 \\ \hline

\multicolumn{9}{c}{$^a$Fan et al. (2003); $^b$Walter et al. (2003); $^b$Bertoldi et al. (2003); $^c$White et al. (2003)} \\

\multicolumn{9}{c}{$^d$Fan et al. (2001); $^e$Fan et al. (2004); $^f$Maiolino et al. (2003); $^g$Iwamuro et al. (2004)}

\end{tabular}
\end{small}
\end{center}
\end{table*}

In this paper we associate the red edge of the GP troughs in the
spectra of $z>6$ quasars with the presence of an \ion{H}{2} region embedded
in an IGM with neutral fraction $x_{\rm HI}$. This assumption will be
discussed further in \S\ref{mfp}.  The key parameters in the analysis
presented herein are therefore the host galaxy redshift and the redshift
for the onset of GP absorption. Considering the host galaxy
redshift, the most accurate redshift estimate comes from the CO line
emission from the host galaxy of J1148+5251, for which the uncertainty is
$\Delta z = \pm 0.001$ (Walter et al. 2003; Bertoldi et al. 2003).  Note
that accurate astrometry has shown that the CO emission is coincident with
the optical QSO to within 0.1$''$, providing strong evidence that the CO
emission is from the QSO host galaxy (Walter et al. 2004). Four other
sources in Table 1 have reasonably accurate \ion{Mg}{2} redshifts. Although
\ion{Mg}{2} is a broad line (few thousand km s$^{-1}$), it is well
documented to provide a relatively accurate estimate (within a few hundred
km s$^{-1}$, or $\Delta z = \pm 0.01$) of the host galaxy redshift in low
redshift quasars (Tytler \& Fan 1992; Richards et al. 2002). This
expectation is supported by the good agreement between the CO and
\ion{Mg}{2} redshifts for J1148+5251. Two of the sources, J1623+3112 and
J1602+4228, have only optical redshifts, corresponding to Ly$\alpha$ and
high ionization broad lines.  The accuracy of the Ly$\alpha$ emission line
redshift is poor, due to strong associated absorption. Similarly, the high
ionization lines are well known to be systematically blueshifted relative
to their host galaxies, with a mean offset of $824 \pm 510$km$\,$s$^{-1}$,
as seen in a large sample of SDSS quasars (Richards et al. 2002). Hence, we
assume a large error of $\Delta z = \pm 0.03$ ($\sim$1200 km$\,$s$^{-1}$) for
these lines. In our analysis we rely on the CO, \ion{Mg}{2}, and optical
line redshifts, in descending order of preference.

We estimate the GP redshift from the published spectra, as the redshift at
which the emission extending blueward of the Ly$\alpha$ (or Ly$\beta$) peak
becomes comparable to the noise.  For one quasar, J1030+0524, the on-set of
the Gunn-Peterson effect is well determined via sensitive Keck spectra of
the Ly$\beta$ region of the spectrum (White et al.~2003). A second source,
J1148+5251, has a sensitive Keck spectrum in the Ly$\alpha$ region (White
et al. 2003), from which the on-set of GP absorption can be estimated
fairly accurately. However, using Ly$\alpha$ has the problem that the
damping wing of a significantly neutral IGM just outside the Stromgren
sphere can extend into the GP region of the spectrum, thereby decreasing
the apparent size of the sphere (Mesinger \& Haiman 2004). For the one well
documented case, J1030+0524, this effect amounts to $\Delta z = 0.02$
(White et al.~2003), and we adopt this as the minimum error for the GP
redshift based only on Ly$\alpha$.  For most of the sources in
Table~\ref{tab1} the on-set of the GP effect must be estimated based on the
discovery spectra from 3.5m telescopes, which acquire only a moderate
signal-to-noise ratio. In these cases we again adopt a conservative GP
redshift error of $\Delta z = 0.03$.

In addition to uncovering the onset of a GP trough, the deep
spectrum of SDSS~1148+5251 shows a peak of transmitted flux which is
observed both in the Ly$\alpha$ and Ly$\beta$ troughs, indicating the
presence of a transparent window in the IGM (White et al.~2003). More
puzzling is the observation of transmission peaks in the Ly$\beta$ trough
that do not have counterparts in the Ly$\alpha$ trough. White et al.~(2004)
interpret this as an indication of the presence of two overlapping windows
of transparency; one in the Ly$\beta$ forest at $z\sim6$, and one in the
Ly$\alpha$ forest at $z\sim5$. Although this scenario is unlikely a-priori,
White et al.~(2004) point out that {\it HST} imaging did not reveal a
small cluster of galaxies at $z\sim5$ as had originally been hypothesized.
 It is still possible that a single galaxy, whose angular position is
coincident with the quasar could generate both the \ion{H}{2} region
required for transparency, as well as the flux in the transmitted peak. Such a
galaxy could also explain the presence of low level continuum in the
Ly$\alpha$ and Ly$\beta$ troughs. However as pointed out by Oh \&
Furlanetto~(2004), transmission of continuum is also detected in the
Ly$\gamma$ trough, which argues against a foreground galactic contribution
which would be subject to a Lyman break. Instead Oh \& Furlanetto~(2004)
argue that transmission of continuum in the Ly$\alpha$, Ly$\beta$ and
Ly$\gamma$ troughs points to a highly ionized IGM at $z\sim6.2$ along the
line of sight to this quasar.

Estimates of the neutral fraction based on transmission in the
Ly$\alpha$ and Ly$\beta$ troughs are made at a redshift in the center
of the trough, while estimates based on the size of the
\ion{H}{2} region are made at the redshift of the quasar. For the
purpose of the present work we assume that the red edge of the
Ly$\alpha$ and Ly$\beta$ troughs in SDSS~1148+5251 mark the boundary of an
\ion{H}{2} region (this point is discussed further in
\S\ref{mfp}). Along any line of sight, the redshift of Ly$\alpha$ transmission will be somewhat
lower than the redshift at which the IGM became reionized. This latter
redshift is expected to have a scatter 
among different lines-of-sight of at least 0.15 (Wyithe
\& Loeb~2004d). It would therefore not be surprising if the IGM were
reionized at $z\sim6.2$ 
along the line of sight to SDSS~1148+5251, but only at $z\sim6$ along the
lines of sight to other quasars.

While better optical and near-IR spectra, as well as more CO host galaxy
redshifts, would improve the analysis presented herein (and such programs
are in progress), we already find that interesting conclusions can be drawn
even for the very conservative errors adopted in Table~\ref{tab1}.  Given
that other uncertainties in the problem (e.g. the ionizing luminosities and
lifetimes of the quasar) are of comparable importance, the precise
determination of redshifts will not affect significantly the statistical
conclusions reached in this study.

\subsection{Could the Red Boundary of the GP Trough Correspond to the Ly$\alpha$ Mean-Free-Path in a Highly Ionized IGM Rather than an \ion{H}{2} Region?}

\label{mfp}

Observations of the Ly$\alpha$ forest show that the number of absorbing
systems increases towards high redshift. This results in a decrease of the
mean-free-path for ionizing photons (Fan et al.~2001; Miralda-Escude 2003)
and an increase in the optical depth for Ly$\alpha$ transmission (i.e. the
GP optical depth) with increasing redshift. Fan et al.~(2001) find that the
mean-free-path for ionizing photons was $\sim7$ physical Mpc at $z\sim5.5$
(comparable to the size of the quasar \ion{H}{2} regions in our
discussion), but had declined below an upper limit of $\sim1$ physical Mpc
within the GP trough along the line of sight to SDSS~1030+0524. In this
context, the IGM transmission just blueward of the Ly$\alpha$ resonance of
the $z\ga 6$ quasars extends out to $\sim5$ physical Mpc away from them,
and must be caused by their ionizing radiation. A natural interpretation is
that these quasars ionize the surrounding IGM to a level well in excess of
that expected from an extrapolation the ionizing background from lower
redshifts. Indeed, the ionizing flux from the quasar SDSS~1030+0524 should
exceed the estimated ionizing background at $z\sim5.5$ out to a distance of
$\sim10$ physical Mpc (at $z\sim6.2$).  These considerations imply that the
mean-free-path of ionizing photons should be at least as large as 5--10
physical Mpc around the $z\ga 6$ quasars.

\begin{table*}[tbp]
\begin{center}
\caption{\label{tab2} Ly$\alpha$ Absorption Properties Near $z>6$ Quasars}
\begin{small}
\begin{tabular}{cccccccccc}
\hline 
\multicolumn{3}{c}{} & \multicolumn{7}{c}{Extrapolated Ly$\alpha$ Optical Depth Surrounding $z>6$ Quasars} \\
$z_{\rm abs}$ & $\tau_{\rm eff}(z_{\rm abs})$ & $J^{\rm b}_{21}(z_{\rm abs})$ & J1148 & J1030 & J1048 & J1623 & J1602 & J1630 & J1306 \\\hline
5$\pm$0.2            &  2.0 $\pm$0.3            &  0.1$\pm$0.02    &  0.5       & 1.9        &    0.1     &  1.1       & 2.2        & 2.1        &  0.9      \\
5.5$\pm$0.2          &  2.5 $\pm$0.3            &  0.1$\pm$0.02    & 0.4        &  1.7       &  0.1       & 1.0        &1.9         &  1.9       & 0.8       \\
\hline
\end{tabular}
\end{small}
\end{center}
\end{table*}

Now suppose the IGM were already highly ionized at $z\ga6$. One may wonder
whether the red edge of the GP trough does not actually
correspond to the boundary of an \ion{H}{2} region, but rather to the
radius where Ly$\alpha$ photons near the quasar are absorbed by residual
\ion{H}{1} in an otherwise highly ionized IGM.  We have argued that in a
highly ionized IGM one could expect the ionized fraction of the IGM within
$\sim10$ Mpc of the quasar SDSS~1030+0524 to exceed that at
$z\sim5.5$. This ionization state would be contrary to the observation of a
GP trough, which points to a rapid change in the ionization
state of the IGM at $z\sim6$ (Fan et al.~2001). Nevertheless, we would like
to critically examine the possibility of a highly ionized IGM and
Ly$\alpha$ absorption from residual neutral hydrogen. To this end we find
whether the ionizing intensity of the quasar at the edge of the HII region
is in excess of that required to lower the GP optical depth
below the observed limit.

Table~\ref{tab2} summarizes the absorber redshift $z_{\rm abs}$
(column~1), effective GP optical depth $\tau_{\rm eff}(z_{\rm abs})$
at $z_{\rm abs}$ (column 2), and background ionizing intensity $J^{\rm
b}_{21}(z_{\rm abs})$ at $z_{\rm abs}$ in units of $10^{-21}~{\rm
erg~s^{-1}~Hz^{-1}~cm^{-2}~sr^{-1}}$ (column 3) as inferred from
spectra of high redshift quasars (Fan et al.~2001).  At a fixed level
of clumpiness in the IGM, the effective GP optical depth $\tau_{\rm
eff}(z_{\rm GP})$ due to neutral hydrogen in ionization equilibrium
with the quasars ionizing radiation field at $z_{\rm GP}$ follows
\begin{equation}
\label{tau}
\tau_{\rm eff}(z_{\rm GP})\sim \tau_{\rm eff}(z_{\rm abs})\left(\frac{1+z_{\rm GP}}{1+z_{\rm abs}}\right)^{9/2}\frac{J^{\rm b}_{21}(z_{\rm abs})}{J^{\rm q}_{21}(R)}
\end{equation}
(see Eq. 6 in Barkana \& Loeb~2004), where $J^{\rm q}_{21}(R)$ is the
quasars ionizing intensity at a distance $R$ from the quasar (column 8
of Table~\ref{tab1}) in units of $10^{-21}~{\rm
erg~s^{-1}~Hz^{-1}~cm^{-2}~sr^{-1}}$. We use a specific ionizing
luminosity of $10^{18.05}$erg$\,$s$^{-1}\,$Hz$^{-1}$ per solar B-band
luminosity (Schirber \& Bullock~2002). These values are listed in
column~9 of table~\ref{tab1}. From equation~(\ref{tau}) we evaluate
the effective optical depth $\tau_{\rm eff}(z_{\rm GP})$ at redshift
$z_{\rm GP}$ for each quasar, extrapolating from absorbers at $z_{\rm
abs}\sim5$ and $z_{\rm abs}\sim5.5$ respectively. The values are
listed in columns 4-10 of table~\ref{tab2} for the $z>6$ quasars
considered in this paper.

The extrapolation of optical depth to redshifts beyond 6 using
equation~(\ref{tau}) assumes a clumping factor that does not evolve in
time. However clumping in the IGM is expected to increase towards low
redshift, as structure grows in the IGM.  For this reason the estimates of
the effective GP optical depths surrounding the high redshift quasars are
larger if extrapolated from absorbers at lower redshift. The most
appropriate estimates of optical depth due to neutral hydrogen in
ionization equilibrium with the quasar flux at a distance $R$ are therefore
given by the second row in Table~\ref{tab2}. These values lie in the range
of $\tau_{\rm eff}=0.1-1.9$, which would allow transmission in the observed
GP troughs out to distances that are well in excess of $\sim 5$Mpc away
from the quasar. If the quasars were emitting into an already ionized IGM,
then the spectra would show the usual proximity effect, with a Ly$\alpha$
forest persisting at redshifts where the GP troughs are actually
observed. We therefore infer that the onset of the GP trough is caused by
diffuse neutral IGM rather than the mean free path of Ly$\alpha$ photons
near the quasar.

\subsection{Could the Ly$\alpha$ Transmission be 
Terminated by a Dense Clump, Rather than the Neutral IGM?}
\label{cloud}

Our analysis of limits on the neutral fraction from the size of \ion{H}{2}
regions relies on the inference that the red end of the GP trough can be
used to estimate the size of \ion{H}{2} regions. However the spectra of the
$z>6$ quasars show the presence of clouds with a substantial column
density. If there had been a cloud of sufficiently high column density to
produce an absorption wing over a sufficiently large wavelength range, the
\ion{H}{2} region would have appeared smaller than it actually is,
resulting in lower limits that are systematically high.  We believe that
this is unlikely for the following reasons.

We start from an observational perspective. The hydrogen column density
required to produce a Ly$\alpha$ optical depth greater than 26 [observed
limits for the best studied example (White et al.~2003)] across a
significant fraction, e.g. 10\%, of the 100\AA ~ within the observed
ionized region is $N = 10^{21}~{\rm cm}^{-2}$ at this redshift. This column
density corresponds to damped Ly$\alpha$ absorbers, which are extremely
rare (Storrie-Lombardi \& Wolfe~2000), $\sim0.1$ per unit redshift between
$z\sim2$ and $z\sim4$. Thus it is very unlikely that a damped Ly$\alpha$
absorber would terminate the GP trough in any one of the 7
quasars, as we are dealing with path lengths of only $\sim5$ Mpc (or a
redshift interval of $\delta z \sim 0.1$).

This small probability can also be substantiated by a simple theoretical
argument.  Our model adopts the distribution of gas clump overdensities
from the numerical simulations of Miralda-Escude et al.~(2000) to derive
the critical overdensity, $\Delta_{\rm crit}$, such that the typical line
of sight to the edge of the \ion{H}{2} region at $\sim5$ Mpc would not
contain any clumps with overdensity larger than $\Delta_{\rm crit}$, while
all clumps with overdensity below $\Delta_{\rm crit}$ are ionized. We find
that $\Delta_{\rm crit}=20$. Barkana \& Loeb (2002) have proven a simple
theorem stating that: {\it ``The fraction of the line-of-sight covered by
gas at a given overdensity is equal to the volume filling fraction of gas
at that overdensity"}. We may therefore find the total column density in
clouds along the line-of-sight with overdensities greater than $\Delta_{\rm
crit}$ as follows. The fraction of mass contained within overdensities
greater than $\Delta_{\rm crit}$ ($\sim 0.07$) is multiplied by the mean
column density of neutral gas along a path through the neutral IGM of
length equal to the size of the \ion{H}{2} region
($1.4\times10^{21}$cm$^{-2}$). We then find the total column density due to
clouds along the line-of-sight with overdensities greater than $\Delta_{\rm
crit}$ to be $0.07\times1.4\times10^{21} = 10^{20}$cm$^{-2}$.  This total
column density falls short by an order of magnitude relative to the value
required to produce a wide wing of absorption that would affect
significantly the size estimate of the \ion{H}{2} region.

\subsection{Distributions of properties among $z>6$ quasars}

\begin{figure*}[hptb]
\epsscale{1.7}  \plotone{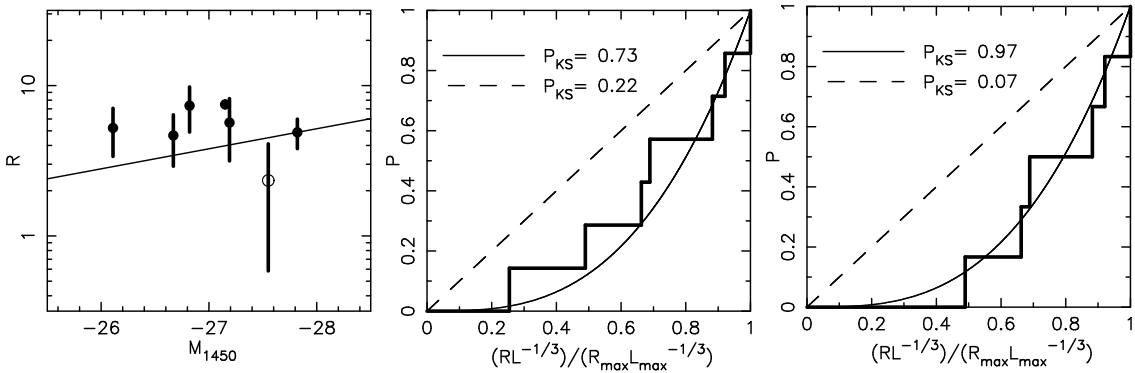}
\caption{\label{fig1} Correlations between the properties of the $z>6$
quasars and their \ion{H}{2} regions. {\em Left panel}: correlation between
quasar absolute magnitude (at rest-frame 1450\AA) ~$M_{1450}$ and inferred
radius of the \ion{H}{2} region around it $R$. The open circle shows the
point for SDSS~1048+4637. The line corresponds to the naive expectation
that $R\propto L^{1/3}$. {\em Central panel}: cumulative probability
histogram for $\eta\equiv RL^{-1/3}/(RL^{-1/3})_{\rm max}$. The solid line
corresponds to $P(<\eta)=\eta^3$, and the dashed line to a random
distribution. {\em Right panel}: same as for the central panel, but with
SDSS~1048+4637 omitted.}
\end{figure*}

We now examine the correlation between the properties of the $z>6$
quasars and the sizes of their \ion{H}{2} regions. The left hand panel
of Figure~\ref{fig1} shows the correlation between the measured
absolute magnitude at a rest-frame wavelength of $1450$\AA,
$M_{1450}$, and the inferred radius of the
\ion{H}{2} region, $R$. The line corresponds to the naive expectation
that $R\propto L^{1/3}$ (normalized to the quasar with the largest
value of $RL^{-1/3}$). Scatter would be expected around this line for
physical reasons (e.g. owing to different quasar ages) in addition to
measurement errors. However, comparison of this line with data points
constitutes an important systematic check that the bubbles do not show
any unexpected trend.

Since the volume within the \ion{H}{2} region grows approximately linearly
with time, the observed bubble sizes should follow a
characteristic distribution.  The central panel of Figure~\ref{fig1}
shows a cumulative probability histogram for $\eta\equiv RL^{-1/3}$
(normalized to the quasar with the largest value of $RL^{-1/3}$ so
that $0\leq \eta \le1$). For a random sample of quasar ages, this
distribution would be expected to follow $P(<\eta)=\eta^3$ (solid
line).  The expected and measured curves are fully consistent in a
Kolmogorov-Smirnov (KS) test with a probability $P_{\rm
KS}\sim0.7$. On the other hand if the observed size of the Ly$\alpha$
transmission region had been set by a dense absorbing cloud rather
than by the neutral IGM, then we would expect a random distribution of
bubble sizes $R$. Since $R$ and $L$ would be uncorrelated in this
case, the distribution of $RL^{-1/3}$ would also be random. However we
find that a random distribution (dashed line) is also consistent in a
KS test, with $P_{\rm KS}\sim0.2$.

We note one possible anomaly in the redshifts listed in
Table~\ref{tab1}. The object SDSS~1048+4637 is the only source with a
value of $z_{\rm MgII}<z_{\rm Ly\alpha,CIV}$, which is not expected
given the distribution of blueshifts for high ionization lines
(Richards et al.~2002). SDSS~1048+4637 is also the only object with a
value of $z_{\rm MgII}-z_{\rm GPT}$, and hence a size for the
\ion{H}{2} region that is consistent with zero (open circle in the
left hand panel of Figure~\ref{fig1}). One would need an
observationally motivated reason to eliminate this quasar from the
sample; nevertheless, it is interesting to repeat the above comparison
of distributions in the absence of SDSS~1048+4637 (right hand panel of
Figure~\ref{fig1}). In this case we see that while the distribution is
consistent with volumes of \ion{H}{2} regions that grow linearly with
time ($P_{\rm KS}\sim0.9$), a random distribution is now only
consistent with the data at the 7\% level.  In the future, using a
larger sample of quasars (perhaps from the completed SDSS) we may be
able to reject the random distribution. For now, we will return to the
possibility that dense absorbing clouds might mimic the signature of
\ion{H}{2} regions in a neutral IGM in section~\ref{cloud}.

Another possible source of systematic uncertainty arises from
gravitational lensing. The large magnification bias for quasars on the
bright end of the quasar luminosity function implies that strong
gravitational lensing may be one or even two orders of magnitude more
common in the SDSS $z\sim6$ quasars than in lower redshift samples
(Wyithe \& Loeb~2002a;2002b; Commerford, Haiman \& Schaye~2002).
Undetected lensing leads to an overestimate of the quasar
luminosity. In the present analysis this leads to an overestimate of
the neutral fraction. Currently there is high resolution imaging for
four of the $z>6$ quasars. As part of an HST snapshot survey of high
redshift quasars, SDSS~1030+0524 and SDSS~1306+0356 have been imaged
(Richards et al.~2003). In addition Keck K-band images have been
obtained for SDSS~1048+4637 and SDSS~1148+5251 (Fan et al.~2003),
while SDSS~1148+5251 has also been imaged with HST (White et
al.~2004). None of these images show evidence for strong lensing, and
hence for significant magnification (however see Keeton, Kuhlun \&
Haiman~2004).  As a result we do not consider lensing in our analysis.

\section{The model}
\label{model}

The model used in this paper to compute the evolution of quasar
\ion{H}{2} regions has been previously described in Wyithe
\& Loeb~(2004a). However for completeness we present a summary of its
main features below.

Quasars are assumed to be powered during the hierarchical growth of their
host galaxy.  Bright episodes are triggered when halos merge. Within our
model, we generate many random realizations of the merger tree of the host
halo at $z=6.5$, assign super-massive black holes (SMBHs) to these halos and
compute the time-dependent luminosity that is triggered during the merges.
Each tree has $N_{\rm merge}$ major mergers, which occur at times $t_j$.
We assume a relation between the black-hole and galaxy halo mass that
preserves the correlation between the circular velocity of the halo and
the black-hole mass it hosts in the form
\begin{equation}
\label{MbhMhalo}
M_{\rm halo} = 1.5\times10^{12}M_\odot\left(\frac{M_{\rm bh}}{10^9M_\odot}\right)^{3/5}\left(\frac{1+z}{7}\right)^{-3/2}.
\end{equation}
Inferred black-hole masses for the SDSS quasars of $M_{\rm bh}\sim
10^9M_\odot$ imply halo masses of $M_{\rm halo}\sim 10^{12}M_\odot$ (Wyithe
\& Loeb 2003). Since this relation is non-linear, there is a mass
differential between the coalesced black-hole mass and the mass of the
new halo. We define a parameter $f_{\rm lt}\equiv f_{\rm
mass}/\eta_{\rm Edd}$, where $f_{\rm mass}$ is the fraction of the
mass differential that is accreted during the luminous phase and
$\eta_{\rm Edd}$ is the fraction of the Eddington rate at which the
mass is accreted. If a merger of two halos leads to coalescence of
their black-holes, and the mass differential is accreted during a
luminous phase over which the quasar shines near its Eddington
limiting luminosity, then the quasar lifetime is
\begin{equation}
\label{lt}
t_{\rm lt} = f_{\rm lt} \times 4\times10^7\left(\frac{\epsilon}{0.1}\right)\ln\left[\frac{\left(M_1+M_2\right)^{5/3}}{M_1^{5/3}+M_2^{5/3}}\right]\hspace{2mm}\mbox{years},
\end{equation}
where $M_1$ and $M_2$ are the masses of the merging halos, and $\epsilon$
(taken to be 0.1) is the radiative efficiency. Note that the lifetime
increases for sub-Eddington accretion.  The parameter $f_{\rm lt}$ may be
thought of as the fraction of our fiducial lifetime during which the
quasars shine, leading to quasar lifetimes of $t_{\rm lt}\sim f_{\rm
lt}\times10^7$ years. The model lifetime is therefore consistent with
current estimates [$10^6-10^8$ years; see Martini~(2003) for a summary] for
values of $f_{\rm lt}$ that are of order unity. We assume that each quasar
episode $j$ has an exponential light-curve
\begin{equation}
\dot{N}_j(t)=\Theta\left(t-t_{j}\right)\dot{N}_{0,j}\exp{\lbrace-(t-t_j)/t_{\rm lt}\rbrace}
\end{equation}
beginning at $t_{j}$ and with a characteristic decay time of $t_{\rm
lt}$ during which the SMBH shines at its Eddington luminosity,
$L_E=1.4\times 10^{47}~{\rm erg~s^{-1}}(M_{\rm bh}/10^9M_\odot)$. Here
$\Theta$ is the Heaviside step function. The time dependent ionizing
luminosity within the merger tree is then computed from the sum of
luminous episodes $\dot{N}(t)=\sum_{j=1}^{N_{\rm
merge}}\dot{N}_j$. Following Telfer et al.~(2002) and White et
al.~(2003), we adopt an ionizing photon emission rate of
$\dot{N}_{0,j}=2\times10^{57}$s$^{-1}/\left({\alpha_{{\rm
EUV},j}}/{-1.75}\right)$, where $\alpha_{{\rm EUV},j}$ is the spectral
index in the EUV band during the $j$th merger. Telfer et al.~(2002)
find $\alpha_{\rm EUV}=-1.75\pm0.75$, and we assign a value of
$\alpha_{{\rm EUV},j}$ to each merger in the tree from a normal
distribution of mean -1.75 and standard deviation 0.75.

The evolution of the physical radius of the \ion{H}{2} region, $R$, may
then be found through integration of the differential equation
\begin{equation}
\label{Vev2}
\frac{dR}{dt}=c\left[\frac{F_\gamma\dot{N}(t) - \alpha_{\rm B}C F_{\rm m} x_{\rm HI}\left(\bar{n}_0^{\rm
H}\right)^2 \left({4\pi\over 3}R^3\right)}
{F_\gamma\dot{N}(t) + 4\pi R^2 \left(1+z\right)^{-1} c F_{\rm m} x_{\rm
HI}\bar{n}_0^{\rm H}}\right],
\end{equation}
where $c$ is the speed of light, $\bar{n}_0^{\rm H}$ is the mean number
density of protons at $z=0$, $\alpha_{\rm
B}=2.6\times10^{-13}$cm$^3$s$^{-1}$ is the case-B recombination coefficient
at the characteristic temperature of $10^4$K, and $\dot{N}_{\rm ion}$ is
the rate of ionizing photons crossing a shell at the radius of the
\ion{H}{2} region at time $t$.  We use the distribution derived from
numerical simulations for the over-densities $\Delta$ in gas clumps
(Miralda-Escude, Haehnelt \& Rees~1998), and calculate the mean free path
for ionizing photons $d(\Delta_{\rm c})$ as a function of the critical
overdensity $\Delta_{\rm c}$ (Miralda-Escude, Haehnelt \& Rees~1998;
Barkana \& Loeb~2002).  Following Barkana \& Loeb~(2002), we then find the
value of $\Delta_{\rm c}$ at which a fraction $\exp\left[-R/d(\Delta_{\rm
c})\right]=50\%$, of the emitted photons do not encounter an overdensity
larger than $\Delta_{\rm c}$ within the \ion{H}{2} region.  We also compute
the mass fraction $F_{\rm m}$ ($\sim1$) of gas within $R$ that is at
over-densities lower than $\Delta_{\rm c}$.  Finally, we calculate the
clumping factor in the ionized regions,
$C(R)\equiv\langle\Delta^2\rangle/\langle\Delta\rangle^2$, where the
angular brackets denote an average over all regions with
$\Delta<\Delta_{\rm c}$. For $x_{\rm HI}=1$ and $dR/dt\ll c$,
equation~(\ref{Vev2}) reduces to its well-known form (e.g. Madau \&
Rees~2001).  The emission rate of ionizing photons $\dot{N}$ in
equation~(\ref{Vev2}) is computed at $t'=t-t_{\rm delay}$ to account for
the finite light travel time between the source and the ionization front.

Note that while equation~(\ref{Vev2}) is expressed in terms of the
radius of a spherical \ion{H}{2} region, there is no implicit
assumption about isotropy in the analysis presented in this paper. This
is because both the volume of the HII region and the luminosity of the
quasar are measured per unit solid angle along the line-of sight. The
extrapolations to total volume and total luminosity are made by
multiplying these quantities by $4\pi$ purely for convenience of
presentation.

In our calculation we require that the quasar ionize regions up to a
sufficiently high overdensity, so as to allow the ionizing photon
mean-free-path to exceed the radius of the \ion{H}{2} region.  The highest
density regions (which are sufficiently compact to allow a long
mean-free-path) may remain neutral.  The value of $x_{\rm HI}$ in
equation~(\ref{Vev2}) refers to the neutral fraction in the low density
regions. We therefore interpret the neutral fraction $x_{\rm HI}$ in
equation~(\ref{Vev2}) as volume weighted rather than mass weighted
fraction. 

In the hierarchical picture of structure formation, the
appearance of the quasar and the surrounding galaxies will occur
concurrently. The neutral fraction into which the quasar \ion{H}{2} region
expands therefore reflects the contribution to reionization due to stellar
flux from the quasar host and surrounding galaxies (see Fig. 2 in Wyithe \&
Loeb 2004c for their relative significance). As the quasar and stellar
ionizing fluxes are emitted at the same cosmic epoch, we implicitly assume
that both are responsible for reionizing the low density regions, such that
both have an ionizing photon mean free path that exceeds the radius of the
\ion{H}{2} region. In this our model differs markedly from the work
described in Yu \& Lu~(2004), where quasar flux is assumed to be emitted
into an IGM whose low density regions have already been ionized by stars at
some prior epoch. In that work the clumping factor associated with quasar
ionizing flux is evaluated {\em above} the density threshold corresponding
to a mean-free-path that equals the size of the \ion{H}{2} region, i.e. it
is assumed that stars ionize the low density regions while the quasar
ionizes only the high density regions. As a result, Yu \& Lu~(2004) infer a
much higher clumping factor than we find here, and conclude in difference
to this work that the quasar flux does not provide a significant
contribution to the growth of the \ion{H}{2} regions.

Several consequences of our simple model can be immediately confronted with
observational data.  Evidence from direct determination of the $M_{\rm
bh}$--velocity dispersion relation (Shields et al.~2003), as well as the
redshift evolution of the quasar correlation function (Wyithe \&
Loeb~2004b), suggest that the assumed relation (equation~\ref{MbhMhalo}) is
indeed preserved out to a redshift of at least $z\sim 3$. Preliminary
support for an extrapolation of the relation out to $z\sim6$ comes from the
velocity shift of the Ly$\alpha$ absorption feature due to the galactic
virialization shock, whose amplitude gauges the circular velocity and hence
mass of the host halo for some high-redshift quasars (Barkana \& Loeb
2003). The presence of a massive galaxy is also implied by the molecular
mass of $>10^{10}M_\odot$ in the host galaxy of SDSS1148$+$5251 (Walter et
al.~2003; 2004) and the velocity width of $280~{\rm km~s^{-1}}$ for its CO
lines (Bertoldi et al. 2003). The measured CO velocity width corresponds to
a dark halo mass of $\sim10^{11}M_\odot$ at $z=6.4$, which is insufficient
to contain the inferred molecular gas mass. Indeed, $280$km$\,$s$^{-1}$ is
much smaller than the $\sim500$km$\,$s$^{-1}$ we would expect to be
associated with a SMBH of $\sim10^9M_\odot$ (Wyithe \& Loeb~2003,2004b),
which is thought to power the observed quasar. A possible explanation for
this inconsistency is that the CO observations only sample the
gravitational potential within a few kpc from the galaxy center while the
galaxy halo has a much deeper potential well.  To quantify this uncertainty
we note that the number of major mergers per logarithm of mass increases by
a factor of $\sim2.4$ if the host dark matter halo mass for the SDSS
quasars is an order of magnitude smaller than assumed
(i.e. $10^{11}M_\odot$ rather than $10^{12}M_\odot$). Thus, adopting a
smaller dark matter halo at the bottom of the merger tree would lead to a
higher merger rate, a more frequent quasar activity, and hence the
inference of a larger neutral fraction.

Additional support for the assumed $M_{\rm halo}-M_{\rm bh}$ relation comes
from attempts to model the luminosity function of quasars using the
abundance of halos and the $M_{\rm bh}-\sigma$ relation (Volenteri, Haardt
\& Madau 2002; Wyithe \& Loeb 2003). These models are equally successful at
$z\sim6$ as they are at lower redshifts $z\sim2$--$3$ where data exists on
the $M_{\rm bh}-\sigma$ relation.  These physically motivated models for
the luminosity function also suggest a quasar lifetime that is in the
vicinity of $10^6-10^7$ years at $z\sim6$. Comparing with $t_{\rm lt}$ (as
calculated from equation~\ref{lt}) implies a value $f_{\rm lt}>0.1$, or in
other words that most of the SMBH mass at $z>6$ was accreted during the
luminous phase. Indeed having all the black-hole mass accreted during the
luminous phase ($f_{\rm lt}=1$) is consistent with the census of mass in
local dormant SMBHs, compared with the mass accreted during luminous quasar
phases throughout the history of the universe. These studies are most
sensitive to conditions at $z\sim2-3$, but find the majority of SMBH mass
to have been accreted during luminous quasar phases near the Eddington
limit, and with a radiative efficiency of $\epsilon\sim0.1$ (Yu \&
Tremaine~2002).

To estimate the ionizing luminosity of the $z>6$ quasars we use the median
spectrum of low redshift quasars derived by Telfer et al.~(2002), scaled to
the appropriate luminosity of the quasar ($M_{1450}$). The assumption of no
evolution of the quasar spectrum over 90\% of the age of the universe is
supported by the observation that the median rest frame UV spectrum of the
high redshift quasars is consistent with that at low redshift (Fan et
al.~2004). In addition, the recently observed X-ray spectrum of
SDSS~1306+0356 implies an optical to X-ray spectral index that is
consistent with radio-quiet quasars at lower redshift (Schwartz \&
Virani~2004). These results imply little evolution in quasar spectra, and
justify our use of the low redshift median quasar spectrum for the analysis
of the $z>6$ quasars.

\begin{figure*}[hptb]
\epsscale{1.7} \plotone{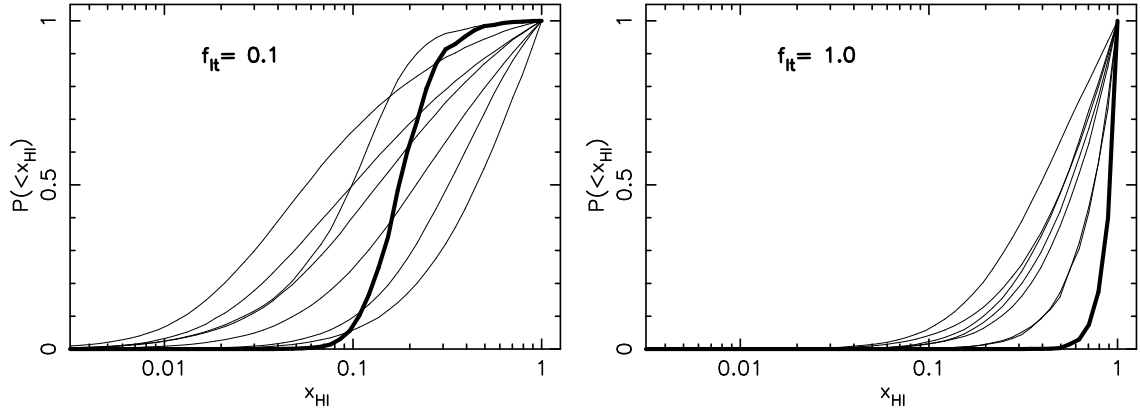}
\caption{\label{fig2} Cumulative probability for the neutral fraction of
the IGM, $x_{\rm HI}$. The thin lines show the distributions for individual
quasars, while the thick line shows the combined result. {\em Left}:
$f_{\rm lt}=0.1$. {\em Right}: $f_{\rm lt}=1$.}
\end{figure*}

\section{Constraints on the Neutral Fraction}
\label{constraints}

Next we derive limits on the neutral fraction of hydrogen surrounding the
$z>6$ quasars. Our basic method follows that outlined in Wyithe \&
Loeb~(2004a). However we have augmented the approach to reflect the richer
data set now available.

\begin{figure*}[hptb]
\epsscale{1.7}  
\plotone{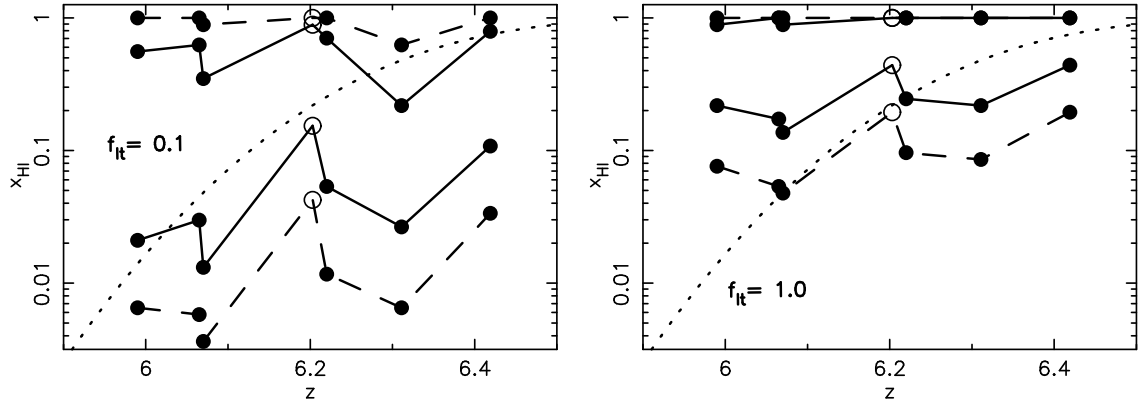}
\caption{\label{fig3} Limits from individual quasars as a function of their
redshift. Each quasar is represented by a point. From bottom up, the
curves show the 1st, 10th, 90th and 99th percentiles of $P(<x_{\rm
HI})$.  The left and right-hand panels show the cases of $f_{\rm
lt}=0.1$ and $f_{\rm lt}=1$, respectively, and limits derived from
SDSS~1048+4637 are denoted by an open circle. The dotted curves
provide the integral of a normal distribution in redshift
$\int_0^zdz'N(6.4,0.15)$.}
\end{figure*}

We compute the conditional probability distributions
$\left.\frac{dP_i}{dR}\right|_{x_{\rm HI}}$ for the observed radius of
the \ion{H}{2} region surrounding each quasar as a function of the
neutral fraction of hydrogen, $x_{\rm HI}$.  Based on these
distributions we find the likelihood for the neutral fraction from the
observed radius around each quasar
\begin{equation}
\nonumber
L_i(x_{\rm HI}) = \int dR_i N(\bar{R}^{\rm obs}_i,\sigma_{\rm R_i})\left.\frac{dP_i}{dR}\right|_{x_{\rm HI}}\hspace{-5mm}(R=R_i)
\end{equation}
Here $N(x,\sigma_x)$ is a normal distribution of mean $x$ and variance
$\sigma_x$. The values of mean and variance of the radii are given
in Table~\ref{tab1}. Assuming the neutral fraction to have the same
value around all quasars in the sample, we also find the joint likelihood
\begin{equation}
L(x_{\rm HI}) = \Pi_{i=1,7}L_i.
\end{equation}

The relative likelihood for $x_{\rm HI}$ may be combined with its a-priori
probability distribution $dP_{\rm prior}/dx_{\rm HI}$ to yield
cumulative a-posteriori probability distributions
\begin{equation}
P_i(<x_{\rm HI})=N_i\int_0^{x_{\rm HI}}dx'_{\rm HI}L_i(x'_{\rm HI})\frac{dP_{\rm prior}}{dx'_{\rm HI}},
\end{equation}
and
\begin{equation}
P(<x_{\rm HI})=N\int_0^{x_{\rm HI}}dx'_{\rm HI}L(x'_{\rm HI})\frac{dP_{\rm prior}}{dx'_{\rm HI}},
\end{equation}
where $N$ and $N_i$ are normalizing constants such that $P_i(<1)=1$
and $P(<1)=1$ respectively. In this paper we adopt a logarithmic prior
for $x_{\rm HI}$, i.e. $dP_{\rm prior}/dx_{\rm HI}\propto 1/x_{\rm
HI}$ for $10^{-3}<x_{\rm HI}<1$, corresponding to the range allowed by
observations of the GP trough (White et al.~2003). More
stringent limits are found for $x_{\rm HI}$ if a flat prior is
assumed.

In Figure~\ref{fig2} we plot the cumulative probability of $x_{\rm
HI}$. The thin lines show the distributions for individual quasars,
whereas the thick line shows the combined result. The left panel shows
the case of $f_{\rm lt}=0.1$, corresponding to the situation where the
quasar fiducial lifetime in equation~(\ref{lt}) overestimates the true
lifetime by a factor of 10. The right hand panel shows the fiducial
case, with $f_{\rm lt}=1$. We note that each of the seven quasars
individually yields a consistent result that $x_{\rm HI}$ is of order
0.1--1. The combined constraint shown by the thick line is therefore
not controlled by observations of just one or a couple of the quasar
\ion{H}{2} regions.  The combined constraint for the fiducial model is
$x_{\rm HI}>0.1$ ($f_{\rm lt}=0.1$) and $x_{\rm HI}>0.7$ ($f_{\rm
lt}=1$) at the 90\% confidence level. Clearly $f_{\rm lt}$ is a
limiting systematic uncertainty and we return to its dependence
below.

In Figure~\ref{fig3} we examine the dependence of the individual
quasar limits on the quasar redshift. Each quasar is represented by 4
points (with SDSS~1048+4637 denoted by open circles), which from
bottom up show the 1st, 10th, 90th and 99th percentiles of $P_i(<x_{\rm
HI})$. Lines join these points for each percentile to guide the
eye. The left and right-hand panels illustrate the cases of $f_{\rm
lt}=0.1$ and $f_{\rm lt}=1$, respectively.  The apparent trend is that
higher limits on $x_{\rm HI}$ are derived for higher redshift
quasars. This trend is to be expected if the quasars are observed near
the end of the reionization process.

The variation of constraints with quasar redshift may be compared with
expectations from theory.  The redshift at which a region of a given size
is reionized is proportional to the linear overdensity on that scale
(Barkana \& Loeb~2003). Along different lines of sight this redshift
has a Gaussian distribution with a variance given by the power
spectrum of primordial fluctuations.  Wyithe \& Loeb~(2004d) have
shown that the combined constraints of finite light travel time and
cosmic variance imply that the scatter in the redshift at which
neutral IGM would be encountered along a random line of sight is
$\Delta z\sim0.15$. This scatter defines a minimum rate over which the
IGM could become reionized. For comparison with the observed limits,
the dashed line in Figure~\ref{fig3} therefore shows the cumulative
distribution corresponding to a Gaussian with variance $0.15$ around a
central redshift of 6.4 (which results in a neutral fraction of
$x_{\rm HI}\sim10^{-3}$ at $z\sim5.85$ where limits exist based on the
optical depth in the GP trough). The limits from individual
quasars imply a somewhat slower evolution in the neutral fraction than
the maximum rate described by the dashed curve, as expected.

\begin{figure*}[hptb]
\epsscale{.9}  \plotone{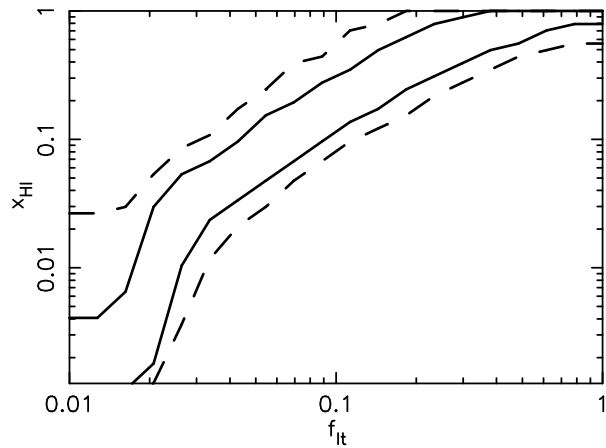}
\caption{\label{fig4} The neutral fraction as a function of $f_{\rm
lt}$. From the bottom up, the lines show the 1st, 10th, 90th and 99th
percentiles of $P(<x_{\rm HI})$ respectively.  }
\end{figure*}

As already mentioned, the largest systematic uncertainty in the analysis
described is the value of $f_{\rm lt}$. While we expect $f_{\rm lt}\sim1$,
it is instructive to compute the limits on the neutral fraction as a
function of $f_{\rm lt}$. In Figure~\ref{fig4} we show the 1st, 10th, 90th
and 99th percentiles of $P(<x_{\rm HI})$ from the bottom up. Smaller values
of $f_{\rm lt}$ yield less stringent limits. However we find that $x_{\rm
HI}>0.1$ for all $f_{\rm lt}>0.1$, and $x_{\rm HI}>0.01$ for all $f_{\rm
lt}>0.025$ at 90\% confidence. These limits represent a significant
improvement upon the limits for the volume averaged neutral fraction based
on the optical depth in the GP trough, $x_{\rm HI}>10^{-3}$
(White et al. 2003).

\section{Conclusion}
\label{conclusion}

In this paper we have extended an earlier analysis of the neutral
fraction of hydrogen in the IGM around SDSS~1148+5251 and
SDSS~1030+0524 (Wyithe \& Loeb~2004a) to include all seven quasars now
known at $z\ga 6$. We have used updated redshifts for the hosts of
these quasars and incorporated uncertainties in the measured size of
their \ion{H}{2} region. The small size of the \ion{H}{2} regions
implies that the typical neutral hydrogen fraction of the IGM beyond
$z\sim6$ is in the range 0.1--1. This result also holds for the IGM
surrounding each individual quasar when the \ion{H}{2} regions are
considered separately. The primary systematic uncertainty in the
analysis is the quasar lifetime. However by combining the limits for
the six quasars, we find that at the 99\% level, the neutral fraction
is larger than 0.08 for quasar lifetimes $>10^6$ years, or larger than
0.6 for lifetimes $>10^7$ years. These lifetimes correspond to only
$10^{-3}$ and $10^{-2}$ Hubble times respectively at $z\sim6$ and are
favored by empirical constraints on the lifetime of lower-redshift
quasars (e.g. Martini 2003). Larger duty-cycles lead to stronger limits. We
find that the size distribution of \ion{H}{2} regions is consistent
with the expected distribution for observations that are made randomly
in time. In addition, we find that constraints on the neutral fraction
obtained from individual quasars are stronger at higher redshift. This
is to be expected if the universe is nearing the end of reionization
at $z\sim6$ as the neutral fraction drops with time. A larger sample
of quasar \ion{H}{2} regions from the full SDSS, or those discovered
by forthcoming redshifted 21cm surveys (Wyithe \& Loeb~2004e) will
allow more stringent checks of these trends.

\acknowledgements 
The authors wish to thank Fabian Walter for helpful comment and
discussion during the course of this investigation. This work was
supported in part by NASA grant NAG 5-13292, and by NSF grants
AST-0071019, AST-0204514 (for A.L.).


\begin{thebibliography}{}


\bibitem[]{}
Barkana, R., Loeb, A. 2001, {Phys. Rep.}, {349}, 125

\bibitem[]{}
---------------------. 2002, ApJ, 578, 1

\bibitem[]{}
---------------------. 2003, Nature, 421, 341

\bibitem[]{}
---------------------. 2004, ApJ, 601, 64 

\bibitem[]{}
Bertoldi, F., et al., 2003, Astron. Astrophys, Lett., 409, L47

\bibitem[]{}
Cen, R. 2003, {ApJ}, {591}, L5

\bibitem[]{} 
Cen, R., Haiman, Z., 200, ApJ, 542, L74

\bibitem[Comerford, Haiman, \& Schaye(2002)]{2002ApJ...580...63C} 
Comerford, J.~M., Haiman, Z., \& Schaye, J.\ 2002, ApJ, 580, 63 

\bibitem{}
Fan, X., et al. 2001  AJ,  122, 2833
 
\bibitem{}
--------------------. 2003, astro-ph/0301135
 
\bibitem[]{}
--------------------. 2004, {AJ}, in press; astro-ph/0405138

\bibitem[]{} 
Furlanetto, S.R. , Hernquist, L. \& Zaldarriaga, M. 2004, MNRAS, submitted, astro-ph/0406131

\bibitem[]{}
Freudling W., Corbin, M.R., Korista, K.T., 2003, ApJ, 587, L67

\bibitem[]{}
Gnedin, N.Y., Prada, F. 2004, ApJ, 608, L77

\bibitem[]{} 
Gunn, J.~E.~\& Peterson, B.~A.\ 1965, ApJ, 142, 1633

\bibitem[]{}
Iwamuro et al. 2004, astroph/0408517

\bibitem[]{560} 
Keeton, C., Kuhlun, M. Haiman, Z., 2004, astro-ph/0405143

\bibitem[]{560} 
Kogut, A. et al. 2003, ApJ, submitted; astro-ph/0302213

\bibitem[]{} 
Madau, P., Rees, M.J., 2000, ApJ, 542, L69

\bibitem[]{} 
Maiolino et al. 2003, astroph/0312402

\bibitem[Miralda-Escud{\' e}, Haehnelt, \& Rees(2000)]{2000ApJ...530....1M}
Miralda-Escud{\' e}, J., Haehnelt, M., \& Rees, M.~J.\ 2000, \apj, 530, 1

\bibitem[]{} 
Martini, P. 2003, to appear in "Carnegie Observatories
Astrophysics Series, Vol. 1: Coevolution of Black Holes and Galaxies,"
ed. L. C. Ho (Cambridge: Cambridge Univ. Press); astro-ph/0304009

\bibitem[]{}
Mesinger, A. \& Haiman, Z. 2004, ApJ, submitted; astro-ph/0406188

\bibitem[]{} 
Miralda-Escude, J. 2003, ApJ, 597, 66

\bibitem[]{} 
Oh, S.P., Furlanetto, S.R., 2004, ApJL, submitted, astro-ph/0411152

\bibitem[]{}
Rhoads, J.E., {et al.}, {ApJ}, in press, astro-ph/0403161

\bibitem[]{} 
Richards, G.T., et al., 2002, Astron. J. Supp., 124, 1

\bibitem[]{} 
--------------------. 2003, Astron.J., 127, 1305

\bibitem[Schirber \& Bullock(2003)]{2003ApJ...584..110S} 
Schirber, M.~\& 
Bullock, J.~S.\ 2003, \apj, 584, 110 

\bibitem[]{}
Schwartz, D. A., \& Virani, S. N.~2004, preprint, astro-ph/0410124

\bibitem[]{}
Shields, G.A., {et al.} 2003, {ApJ} {583}, 124

\bibitem[]{}
Spergel, D. N, {et al.} 2003, {AJ Supp.}, {148}, 175

\bibitem[]{} Storrie-Lombardi, L. J. \& Wolfe, A. M. ~2000, ApJ, 543, 552;
Erratum-ibid. 592, 1263

\bibitem[]{}
Telfer, R. C., Zheng, W., Kriss, G. A., Davidsen, A. F. 2002, {AJ}, {565}, 773

\bibitem[Tytler \& Fan(1992)]{1992ApJS...79....1T} 
Tytler, D.~\& Fan, X.\ 
1992, \apjs, 79, 1 

\bibitem[]{}
Volonteri, M., Haardt, F., Madau, P., 2003, ApJ, 582, 559

\bibitem[]{}
Walter, F., et al., 2003, Nature, 424, 406

\bibitem[]{}
Walter, F., Carilli, C., Bertoldi, F., Menten, K., Cox, P., Lo, K.Y., Fan, X., Strauss, M., 2004, ApJL, accepted, astro-ph/0410229

\bibitem[]{}
White, R., Becker, R., Fan, X., Strauss, M. 2003, {AJ}, {126}, 1 

\bibitem[]{}
White, R., Becker, R., Fan, X., Strauss, M. 2004, astro-ph/0411195 

\bibitem[Willott, McLure, \& Jarvis(2003)]{2003astro.ph..3062W} 
Willott, C.~J., McLure, R.~J., \& Jarvis, M.~J.\ 2003, astro-ph/0303062

\bibitem[Wyithe \& Loeb(2002)]{2002Natur.417..923W} 
Wyithe, J.~S.~B.~\& Loeb, A.\ 2002a, Nature, 417, 923 

\bibitem[Wyithe \& Loeb(2002)]{2002ApJ...577...57W} 
--------------------. 2002b, ApJ, 577, 57 

\bibitem[]{}
--------------------. 2003a, {ApJ}, {588}, 69

\bibitem[]{}
--------------------. 2003b, {ApJ} {595}, 614

\bibitem[]{}
--------------------. 2004a, {Nature}, {427}, 815

\bibitem[]{}
--------------------. 2004b, {ApJ}, accepted, astro-ph/0403614

\bibitem[]{}
--------------------. 2004c, {ApJ}, submitted, astro-ph/0407162

\bibitem[]{}
--------------------. 2004d, {Nature}, in press, astro-ph/0409412

\bibitem[]{}
--------------------. 2004e, {ApJ}, accepted, astro-ph/0401554

\bibitem[]{} 
Yu, Q., Tremaine, S., 2002, MNRAS, 335, 965

\bibitem[]{} 
Yu, Q., Lu, Y., 2004, ApJ., accepted, astro-ph/0411098


\end{thebibliography}
\end{document}